\documentstyle[pra,eqsecnum,aps,amsmath,amssymb,psfrag,graphicx]{revtex}
\def\>{\rangle}
\def\<{\langle}
\def\n{\nonumber}
\def\lb{\left[}
\def\rb{\right]}
\def\lk{\left\{}
\def\rk{\right\}}
\def\<{\langle}
\def\>{\rangle}
\def\be#1\ee{\begin{equation}#1\end{equation}}
\def\ba{\begin{eqnarray}}
\def\ea{\end{eqnarray}}

\input{psfig}

\begin{document}
\draft
\twocolumn

\title{ Quantum Approach to a Derivation of the
Second Law of Thermodynamics}
\author{Jochen Gemmer, Alexander Otte and G\"unter Mahler}
\address{Institut f\"ur Theoretische Physik,\\Universit\"at Stuttgart, Pfaffenwaldring 57,\\70550 Stuttgart, Germany}
\maketitle


\begin{abstract}
We re-interprete the microcanonical conditions in the quantum domain as
constraints for the interaction of the ``gas-subsystem'' under consideration and
its environment (``container''). The time-average of a purity-measure is found
to equal the average over the respective path in Hilbert-space.
We then show that for typical (degenerate or
non-degenerate) thermodynamical systems almost all states within the allowed
region of Hilbert-space have a local von Neumann-entropy $S$ close to the maximum and 
a purity $P$ close to its minimum, respectively. Typically thermodynamical systems should therefore
obey the second law.
\end{abstract}

\narrowtext
The second law has been formulated in a number of
different ways \cite{TER91}: According to Clausius' postulate heat cannot flow 
spontaneously
from a colder to a hotter system. Thomson's formulation reads: ``It is
impossible to construct a perpetuum mobile of the second kind.'' 
Such rules define, essentially, some sort of irreversibility, i.e. the 
existence of a state
function usually called entropy, $S$, which can only increase in closed systems.\\
The second law is arguably one of the most fundamental
and far-reaching laws of physics; nevertheless its origin remains 
puzzling.\\
On the macroscopic level the second law enables us to calculate state
equations from the requirement that entropy has to be maximized under the
condition of given extensive variables in order to get a thermodynamical
potential as a function of those.\\
On the microscopic level irreversibility comes in conflict with the
notorious reversibility of all fundamental physical laws. It has been the
challenge for statistical mechanics to reconcile $dS/dt \ge 0 $ with the 
underlying
microscopic dynamics.\\
Irreversibility in classical mechanics is conventionally introduced via 
two different schemes (cf. \cite{NEU30}): The Boltzmann approach, based on the 
hypothesis of 
molecular chaos ({\em Stosszahlansatz}) and the Gibbs ensemble approach, 
based on 
the hypothesis of quasi-ergodicity. Both these attempts have to acknowledge some additional
assumptions which do not follow from the underlying microscopic laws. \\
This fact has led to continuing controversies as to whether this situation 
can be the last word. Several researchers suggested quantum mechanics as a
possible remedy.\\
L. D. Landau and E. M. Lifshitz \cite{LAN78} expected that quantum 
measurements -- making a difference between future and past -- should be
responsible for the entropy increase according to the second law. 
This assertion has never been proved, though.\\
E. Schr\"odinger \cite{SCH89} argued that the conventional canonical distributions of
thermodynamics can be obtained also for a system 
described by a single wavefunction, if one invokes 
``the old crux of molecular disorder''.\\
J. von  Neumann \cite{NEU30} was able to show that a non-degenerate system should,
indeed, obey a quantum-mechanical version of ergodicity under fairly weak
conditions. In addition he felt obliged to introduce ``macroscopic observers''
(coarse graining) in order to come into contact with standard thermodynamics.\\
This latter aspect has then been taken up also by W. Pauli and M. Fierz
\cite{PAU37}. These authors insisted that the 2nd law should be explained without
reference to external perturbations like those induced by quantum 
measurements.\\ 
G. Lindblad \cite{LIN83} observed that the entropy of a multi-partite system 
defined as the 
sum of the respective partial entropies of the subsystems would tend to 
increase due to the neglect of 
correlations (entanglement). This fact could be taken as a quantum mechanical
justification of Boltzmann's {\em Stosszahlansatz}.\\  
W. Zurek \cite{ZUR91} and his coworkes have discussed in great detail how the
interaction of a quantum system with its environment may induce quasi-classical
behavior in the former (``environment-induced superselection rules''). 
They argued that the second law should eventually result from the impossibility
of isolating macroscopic systems from their environment \cite{ZUR94}.\\
Many of these ideas are pertinent to our present investigation.
It is the purpose of this Letter to show that under (appropriately
redefined) microcanonical conditions the
second law of thermodynamics follows from a quantum mechanical analysis of the 
total system partitioned into the object and the 
environment.\\
A purity measure that is formulated within standard quantum mechanics but
has also an interpretation within thermodynamics is the
von Neumann-entropy
\be
S(\hat{\rho}):=-k\mbox{Tr}\lk\hat{\rho}\ln \hat{\rho}\rk.
\ee
If this entropy is zero, the system is in a pure state, if it takes on its
maximum value $S_{max}=k\ln(N)$, where $N$ is the system size (number of
accessible states), the system is in
the maximally mixed state.
Another measure is the ``purity'' $P$:
\be
P(\hat{\rho}):=\mbox{Tr}\lk\hat{\rho}^2\rk.
\ee
If $P$ takes on its maximum value $1$ the system is in a pure state, if
$P$ takes on its minimum value $\frac{1}{N}$ the system is in the
maximally mixed state.\\
 For those extreme cases the two measures uniquely map onto each other. For
 general cases, however, this does not hold true anymore. Nevertheless, states with
$P(\hat{\rho})\approx 1$ will have $S(\hat{\rho})\approx 0$ and states with
$P(\hat{\rho})\approx P_{min}$ will have $S(\hat{\rho})\approx S_{max}$. In
the following  we mainly consider $P$ and get back to $S$ in the end.\\
Using the von Neumann-equation for the density operator of the total
system it can easily be shown that the entropy and the purity of a closed system that does not
interact with any other system are conserved.\\
This fact might be considered a contradiction to the second law which
demands that entropy should be maximized during the evolution of a closed
system.\\
If the full system is being
regarded as divided into two subsystems (I and II), the reduced density
operators are:
\be
\hat{\rho}_I=\mbox{Tr}_{II}\lk\hat{\rho}\rk,\quad
\hat{\rho}_{II}=\mbox{Tr}_I\lk\hat{\rho}\rk
\ee
Entropy and purity may be defined for each subsystem as before, using
$\hat{\rho}_I\left(\hat{\rho}_{II}\right)$ instead of $\hat{\rho}$.\\
Since those are not the density operators that appear in the
von Neumann-equation, $S_I\left(S_{II}\right)$ and $P_I\left(P_{II}\right)$ defined on basis of these operators
are no longer conserved. Those are the quantities we are going to examine.\\
Although the following ideas apply to all sorts of subsystems, we want
to refer to the system of which the entropy is to be calculated as the
``gas-system'', g, and all the surrounding as the ``container-system'', c.\\
The full Hamiltonian is now divided according to
the same scheme:
\be 
\hat{H}=:\hat{L}_g+\hat{L}_c+\hat{W}
\ee
$\hat{L}_g$ describes the energies arising from the gas particles alone, including
their mutual interactions. $\hat{L}_c$ describes the corresponding energies
of the container particles
alone. $\hat{W}$ describes the interaction terms that depend on both, the coordinates of gas
particles and the container particles. These are here basically the ``wall''
interactions that keep the gas particles inside the container.\\
The energy eigenstates of a free gas are unbound and continuous. It is far more convenient to have
bound states for the ``separate'' systems. Thus we modify the Hamiltonian formally
in the following way:
\be 
\hat{H}=:\hat{L}_g'+\hat{L}_c+\hat{W}'
\ee
with
\be
\hat{L}_g':=\hat{L}_g+\hat{V} \quad \mbox{and} \quad \hat{W}':=\hat{W}-\hat{V}
\ee
where $\hat{V}$ models the mean effect of the container on the gas
particles. $\hat{V}$ is an effective potential that only depends on the
coordinates of the gas particles and is chosen to minimize $\hat{W}'$. Usually the container is
simply modelled by some ``box'' potential $\hat{V}$, neglecting $\hat{W}'$
altogether. But however small, starting from first principles $\hat{W}'$
will always be present, and represents a coupling.\\
We consider the gas-system to be closed in the thermodynamical sense,
i. e., controlled by microcanonical conditions ($E,V,N=const.$); this is
clearly an idealization but can routinely be realized in an approximate
way.\\
A system that is closed on the macroscopic level (thermodynamically
closed) does not need to be closed on the microscopic (quantum-) level
(i. e. not
interacting with any other system).\\
The fact that no extensive quantities are to be exchanged, however, puts constraints on its Hamiltonian, especially on the interactions a system can
have with its surrounding, in order for it to be thermodynamically
closed.\\
The strict conservation of particles $N$ is taken into account simply by
the way the system is partitioned. To which accuracy the volume $V$ stays fixed
is set by $\hat{L}_c$. Microcanonical conditions then correspond to a box with very high, in the limit of the
volume $V$ being exactly conserved, infinitely high potential walls.\\
The condition that no energy is to be exchanged further constrains the
Hamiltonian. The energy contained in the gas is given by:
\be
E_g:=\<\hat{L}_g'\>.
\ee
If this is to be conserved, it follows that
\be
\label{kommu}
\lb\hat{L}_g',\hat{H}\rb=0 \quad \lb\hat{L}_g',\hat{W}'\rb=0.
\ee
Except for these constraints we need not specify $\hat{W}'$ in more detail.\\
Based on these commutator relations we find that for any energy eigenspace $A,B$
\be
\label{constr}
\sum_{i,j}|\psi_{ij}^{AB}(t)|^2=\sum_{i,j}|\psi_{ij}^{AB}(0)|^2
\ee
is a conserved quantity, set by the initial state, where $\psi_{ij}^{AB}$
denotes the amplitudes of the degenerate product energy eigenstates of
$\hat{L}_g'+\hat{L}_c$ (``$i$'' denoting the gas, ``$j$'' the container
part of the product) that are associated with the energy eigenvalues $E_A^g\left(E_B^c\right)$ in the gas-(container-)
system.\\
Since we want to consider cases here that have zero local entropy in the
beginning (product states), we get
\be
\sum_{i,j}|\psi_{ij}^{AB}(0)|^2=\sum_{i,j}|\psi_i^A(0)|^2|\psi_j^B(0)|^2=P_A^gP_B^c
\ee
where $P_A^g\left(P_B^c\right)$ are the
probabilities of finding the gas-(container-) sytem somewhere in the possibly highly
degenerate subspace characterized by the energy
eigenvalues $E_A^g\left(E_B^c\right)$. If no energy is to be exchanged,
clearly these probabilities have to remain conserved. This is the constraint that
microcanonical conditions impose on the accessible region of Hilbert-space.\\
Although we are interested in $P(t\rightarrow \infty)$, we start by
considering the time average of the purity $P$ for reasons that will become
clear later.
\be
\label{taver}
\overline{P}:=\frac{1}{T}\int_0^T P(|\psi(t)\>)dt
\ee
Choosing a special parametrization for $|\psi\>$, we can convert the time
integral into an integral over the trajectory generated by the total
system's dynamics for given initial conditions. Parametrizing $|\psi\>$ in terms of the real and imaginary parts of
its amplitudes
\be
|\psi(t)\>:=\lk \psi_{ij}(t),\psi'_{ij}(t)\rk
\ee
we can write instead of (\ref{taver})
\be
\label{killer}
\overline{P}=\frac{\int_{|\psi(0)\>}^{|\psi(T)\>}P\left(\lk
  \psi_{ij},\psi'_{ij}\rk\right)\frac{1}{v_{eff}} |d|\psi\>|}{\int_{|\psi(0)\>}^{|\psi(T)\>}\frac{1}{v_{eff}}|d|\psi\>|}
\ee
where $|d|\psi\>|$ denotes the ``length'' of an infinitesimal step along the
trajectory in Hilbert-space.\\
The advantage of this special
parametrization derives from the fact that the effective velocity
\be
v^2_{eff}=\sum_{i,j}\left(\dot{\psi_{ij}}^2+\dot{\psi'_{ij}}^2\right)=\frac{1}{\hbar^2}\<\psi (0)|\hat{H}^2|\psi (0)\>
\ee
is constant on each trajectory and thus independent of the
time $t$ or the special point on the trajectory. Hence, the integral
  (\ref{killer}) simplifies to
\be
\overline{P}=\frac{1}{L}\int_{|\psi(0)\>}^{|\psi(T)\>}P\left(\lk
  \psi_{ij},\psi'_{ij}\rk\right)|d|\psi\>|
\ee
where $L$ is the length of the path. So, the time average of $P$ equals the path average along a special
  trajectory in this parametrization of Hilbert-space.\\ 
We are not able to compute this
  integral for we do not know $\hat{W}'$ in detail, and even if we did, we could never hope to solve the Schr\"odinger-equation for a
  system with about $10^{23}$ degrees of freedom.\\
All we want to prove here, is that for typical trajectories staying within the region allowed by the microcanonical
  conditions, $P$ is extremely close to its minimum value for almost all points within this region.\\
We proceed as follows:\\
First we calculate $P_{min}$ which is the smallest possible value
of $P$ within the allowed region. Then we compute the average of $P$
over the total allowed region. If this average is close to $P_{min}$, we
can conclude that $P\approx P_{min}$ for almost all points within this
region, which means for almost all $P(t)$, since any distribution with a mean value close to a boundary has to
be sharply peaked.\\
From (\ref{kommu}) it follows that the $P_A^g$ remain conserved. Now the
lowest purity $P$ of any state consistent with this conditon is:
\be
\label{pmin}
P_{min}=\sum_A\frac{(P_A^g)^2}{N_A^g}
\ee
where $N_A^g$ is the degree of degeneracy of $E_A^g$.\\
To calculate the Hilbert-space average of $P$ denoted as $<P>$ we
need a parametrization for $\psi_{ij},{\psi'}_{ij}$ confined to the allowed
region (\ref{constr}) that essentially consists of hyperspheres in different parts of the
Hilbert-space of the total system. The Hilbert-space average can then be written as
\be
<P>=\frac{\int P\left(\lk\psi_{ij}(\lk\phi_n\rk),{\psi'}_{ij}(\lk\phi_n\rk)\rk\right)\det{\mathcal F}\prod_n d\phi_n}{\int\det{\mathcal F}\prod_n d\phi_n}
\ee
where $\phi_n$ is the respective set of parameters and ${\mathcal F}$ is the corresponding
functional matrix.\\
This integral can actually be solved analytically. The  techniques are
essentially the same as those used to calculate surface areas of
hyperspheres in the classical statistical analysis of the ideal gas. Since this calculation is rather elaborate we do not want to present it in detail here, but give
  and discuss the result:
\ba
\label{exact}
&&<P>=\\
&&\sum_A\frac{(P_A^g)^2}{N_A^g}\left(1-\sum_B(P_B^c)^2\right)+\sum_B\frac{(P_B^c)^2}{N_B^c}\left(1-\sum_A(P_A^g)^2\right)\n\\
&&+\sum_{A,B}\frac{(P_A^g)^2(P_B^c)^2(N_A^g+N_B^c)}{N_A^gN_B^c+1}\n
\ea
Here $N_B^g$ is the degree of degeneracy of the energy eigenvalue
  $E_B^g$.\\
If the degeneracy of the occupied energy levels is large
enough so that
\be
\frac{1}{N_A^gN_B^c+1}\approx\frac{1}{N_A^cN_B^c}
\ee
which should hold true for typical thermodynamical systems,
(\ref{exact}) reduces to
\be
<P>\approx\sum_A\frac{(P_A^g)^2}{N_A^g}+\sum_B\frac{(P_B^c)^2}{N_B^c}
\ee
The first sum in this expression is obviously exactly $P_{min}$
(\ref{pmin}), so that for systems and initial conditions in which the
second sum is small the allowed region almost only consists of states for
which $P\approx P_{min}$. The second sum will be small if the container
system occupies highly degenerate states, typical for thermodynamical systems.\\
To illustrate this result, we have plotted the relative frequency of $P$
(see Fig 1.), which
we calculated using a formula by Lloyd, Pagels \cite{LLO88} and
Page \cite{PAG93}, that
applies to completely degenerate subsystems only. In this case we find from (\ref{exact})
\be
<P>=\frac{N^g+N^c}{N^gN^c+1}
\ee
(For this special case the average has
also been calculated by Lubkin \cite{LUB78}.)\\
\begin{figure}[!h]
\centering
\psfrag{x}[][]{$P$}
\psfrag{y}[][l]{$P_P$}
\psfrag{p1}[][]{2x2}
\psfrag{p2}[][]{2x4}
\psfrag{p3}[][]{2x8}
\psfrag{0}[][cb]{0}
\psfrag{0.25}[][cb]{}
\psfrag{0.5}[][cb]{$\frac{1}{n_1}$}
\psfrag{0.75}[][cb]{}
\psfrag{1.}[][cb]{1}
\includegraphics[width=6.5cm]{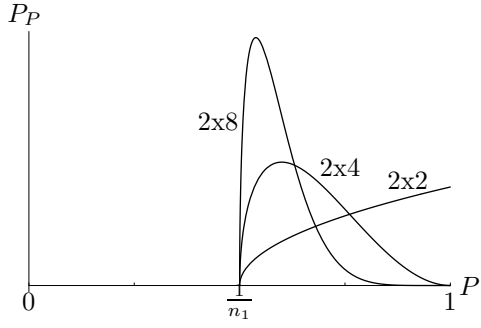}
\vspace{5mm}
\caption{Probability density as a function of $P=\mbox{Tr} \{
  {\hat{\rho}_I}^2 \}$ for
various $n_1$x$n_2$ systems. With increasing size of subsystem II the
density becomes peaked near the value $\frac{1}{n_1}$.}
\label{puritydist}
\end{figure}
Obviously our average $<P>$ is in perfect agreement with the more detailed
distribution of $P$.\\
So far we have only shown that, except for negligible parts, the Hilbert-space-section in which the trajectories can be found under microcanonical
conditions, has $P\approx P_{min}$, provided the surrounding system is much larger than
the considered system.\\
To examine under what conditions trajectories will even
``fill'' the whole space they can possibly live in, we consider the special
case of no degeneracy in either subsystem. Here  the evolution
of $P$ can be calculated exactly:
\be
\label{deco}
P(t)=\sum_{A,B,C,D}e^{\left(\frac{1}{i\hbar}\left(E_{AB}-E_{CB}+E_{CD}-E_{AD}\right)t\right)}P_A^gP_B^cP_C^gP_D^c
\ee
where $E_{IJ}$ are the energy eigenvalues of the respective energy
eigenstates. Note that without any interaction $E_{IJ}=E_I+E_J$ and
$P(t)=const.$, as expected.\\ 
Assuming that all terms in (\ref{deco}) are oscillating except the ones
with $A=C$ or $B=D$, we get
\be
\overline{P}=\sum_A(P_A^g)^2+\sum_B(P_B^c)^2-\left(\sum_A(P_A^g)^2\right)\left(\sum_B(P_B^c)^2\right)
\ee
which is exactly the same result as (\ref{exact}) with $N_A^g=N_B^c=1$. So, under this assumption the Hilbert-space average is exactly equal
to the time average.\\
In general, however, this specific ergodicity is not needed: It suffices
that typical quantum trajectories, even though starting with $P(0)=1$,
venture out into the vast Hilbert-space regions characterized by
$P=P_{min}$. Of course, one cannot exclude that in special situations there might be
trajectories that never leave the very tiny region with $P\approx 1$, but
these situations become extremely rare as the surrounding gets big.\\
Finally, we return to the local entropy $S$. Trying to compute $<S>$ rather than $<P>$ we get, after some lengthy but
straightforward perturbative calculations 
\be
\label{ende}
<S>\approx S_{max}\left(\lk P_A^g,N_A^g\rk
\right)-K\left(\sum_B\frac{(P_B^c)^2}{N_B^c}\right)
\ee
where $K$ is a
 positive function that scales linearly with the system size of the gas
 system. (\ref{ende}) is valid for situations with
\be
\sum_A\frac{(P_A^g)^2}{N_A^g}\gg\sum_B\frac{(P_B^c)^2}{N_B^c},
\ee
which is the thermodynamical regime in which the second term in
(\ref{ende}) will be small.
(Again, for the special case of both subsystems being completely
 degenerate, our results are in perfect agreement with a result by S. Sen \cite{SEN96})\\
In conclusion we have shown that the local von Neumann-entropy of a
considered system will be maximized during its evolution, even if the
system is thermodynamically closed, provided the energy
eigenspaces occupied by the surrounding are much bigger than the energy
eigenspaces occupied by the considered system. This is typical for
thermodynamical systems. We did not need the additional assumptions
underlying classical derivations. Since we considered microcanonical conditions we
get the maximum entropy as an explicit function of  the initial energy distribution. This allows for a connection with standard thermodynamics,
entropy being a thermodynamical potential.\\
Generalizations to canonical conditions and the treatment of quantum
computer systems as specific open systems are under way.\\
We thank Dr. I.~Kim, M.~Stollsteimer, Dipl. Phys. F.~Tonner and T.~Wahl for
fruitful discussions. One of us (A.O.) acknowledeges financial support by
the Deutsche Forschungsgemeinschaft.

\end{document}